\documentclass[sigconf]{acmart}
\usepackage{multirow}

\newcommand{\INIT}{INIT}
\newcommand{\COND}{COND}
\newcommand{\UPD}{UPD}
\newcommand{\DIFF}{Diff}
\AtBeginDocument{%
  }

 \copyrightyear{2026}
 \acmYear{2026}
 \setcopyright{cc}
 \setcctype{by}
 \acmConference[ETRA '26]{2026 Symposium on Eye Tracking Research and Applications}{June 01--04, 2026}{Marrakesh, Morocco}
 \acmBooktitle{2026 Symposium on Eye Tracking Research and Applications (ETRA '26), June 01--04, 2026, Marrakesh, Morocco}
 \acmDOI{10.1145/3797246.3806012}
 \acmISBN{979-8-4007-2519-7/2026/06}

\begin{document}

\title{Identifying Effective Program Comprehension Strategies through Gaze Transitions over Syntactic Elements}

\author{Kyogo Horikawa}
%\authornote{aaaaaaaaaaaaaaaaaaaaaaaaaa}
\email{ai1221@nara.kosen-ac.jp}
\orcid{0009-0002-5930-0166}
\affiliation{%
  \institution{National Institute of Technology(KOSEN), Nara College}
  \city{Yamatokoriyama}
  \state{Nara}
  \country{Japan}
}
\author{Hidetake Uwano}
\email{uwano@info.nara-k.ac.jp}
\orcid{0009-0006-6920-6244}
\affiliation{%
  \institution{National Institute of Technology(KOSEN), Nara College}
  \city{Yamatokoriyama}
  \state{Nara}
  \country{Japan}
}
\author{Haruhiko Yoshioka}
%\authornotemark[2]
\email{yoshioka.haruhiko.yi4@naist.ac.jp}
\orcid{0009-0004-2353-1397}
\affiliation{%
  \institution{Nara Institute of Science and Technology}
  \city{Ikoma}
  \state{Nara}
  \country{Japan}
}

% Additional stuff
%\author{Lars Th{\o}rv{\"a}ld}
%\affiliation{%
%  \institution{The Th{\o}rv{\"a}ld Group}
%\email{larst@affiliation.org}


\renewcommand{\shortauthors}{Horikawa et al.}

\begin{abstract}
Program comprehension is a central research topic in software engineering, focusing on how developers understand a program's structure, behavior, and intent.
Eye-tracking studies have traditionally relied on display-based measurements, where gaze positions are represented as screen coordinates. However, syntax-based analyses have recently emerged.
Prior work proposed methods to convert eye movements into transitions between nodes in an abstract syntax tree, but the relationship between task correctness and eye-movement features for specific syntactic elements remains unclear.
This study converts eye-tracking data into transitions between syntactic nodes and analyzes fixation proportions and gaze transition patterns.
We investigate the relationship between these patterns and task correctness, comparing correct and incorrect groups.
Our results reveal distinct differences in gaze transition patterns between the two groups.
In particular, successful participants exhibit more systematic transitions across syntactic elements, suggesting the use of structured reading strategies.

\end{abstract}



%\keywords{Program comprehension, Eye tracking analysis, Syntactic elements, Pattern mining}



\maketitle

\section{Introduction}

Program comprehension is a research area within software engineering that investigates how developers understand source code.
It focuses on the process by which developers read source code to grasp a program's structure, behavior, and intent.
Analyzing program comprehension helps reveal developers’ cognitive processes and identify challenging program constructs, thereby supporting improvements in code readability and developer education.
Various approaches have been used to study program comprehension, including the measurement of physiological signals, such as eye movements \cite{Busjahn2015} , brain activity \cite{Siegmund2014}, and the analysis of behavioral logs \cite{Ko2006}.
In particular, eye-tracking studies have examined the characteristics of developers’ gaze behavior while reading source code to better understand comprehension processes \cite{ishida2019, sharif2020, Abid2019, Lin2016}.

Most eye-tracking studies in program comprehension have relied on display-positional information, where gaze positions are represented as coordinates on the screen. 
Recently, syntax-based measures have emerged.
Yoshioka proposed a method for converting eye movement data that was recorded  as display coordinates into transitions between nodes in a syntax tree generated from the source code \cite{yoshioka2024}.
The study analyzed the relationship between fixation proportions on different types of syntactic elements and the success or failure of comprehension tasks.
Nevertheless, the relationship between task correctness and eye-movement features for specific syntactic elements remains unclear. 

This study aims to clarify the relationship between program comprehension success and gaze transitions over syntactic elements in source code.
Because program comprehension requires understanding the syntactic structures that determine program behavior, syntactic elements provide semantically meaningful units for analyzing programmers’ code-reading behavior.
Using the coordinate-to-syntax mapping method \cite{yoshioka2024}, we convert gaze data collected during program comprehension tasks into transitions between syntactic nodes. 
We then analyze fixation proportions for each syntactic element and examine how gaze transition patterns differ between successful and unsuccessful participants.
In this article, we investigate the following research questions:

\begin{enumerate}
    \renewcommand{\labelenumi}{RQ\arabic{enumi}}
    \item Is there a difference in gaze transition patterns related to \texttt{for}-loop structures between successful and unsuccessful participants?
    \item Is there a difference in gaze transition patterns across methods between successful and unsuccessful participants?
\end{enumerate}

In this study, we focus on specific syntactic structures to analyze gaze transitions.
Control flow determines the execution order of a program and plays a critical role in program comprehension and task success.
Among control-flow constructs, we focus on the \texttt{for} statement because it frequently appears in the tasks and contains multiple syntactic components (initialization, condition, and update), enabling the analysis of gaze transitions within a single control structure and facilitating the interpretation of transitions between its components.
Also, we focus on methods as representative high-level syntactic elements to analyze gaze transitions.
Methods encapsulate specific functionality within a program and form important structural units for understanding program behavior.
Because program comprehension often involves following execution flow and relationships among different parts of a program, analyzing gaze transitions between methods may reveal how participants track these relationships during comprehension.
Based on this assumption, we hypothesize that gaze transition patterns between methods differ between successful and unsuccessful participants.

%Chapter Structure
The remainder of this paper is organized as follows.
Section 2 reviews related work.
Section 3 describes the coordinate-AST node-mapping method.
Section 4 presents the experimental design and analysis methodology.
Section 5 reports the results and discusses gaze transition patterns.
Finally, section 6 concludes the paper and outlines future work.

%------------------------------------------------------------------------
\section{Related Work}
Yoshioka and Uwano proposed a method for converting gaze movements recorded in display coordinate units by an eye-tracking device into transitions between nodes in a syntax tree generated from source code \cite{yoshioka2024}.
Their study analyzed the relationship between fixation proportions on different syntactic element types and task success, revealing that successful tasks exhibited significantly more fixations on \texttt{if} conditionals and significantly fewer fixations on method formal parameters and \texttt{print} statements.
Furthermore, in complex tasks involving recursion, participants fixated more frequently on statements within \texttt{if} blocks than on the corresponding conditional expressions.
However, while their study focused on fixation proportions for individual syntactic elements, it did not examine gaze transition patterns between syntactic elements or how such transitions relate to task correctness.
The present study extends the research by analyzing both fixation distributions and gaze transitions across syntactic elements, with the aim of clarifying effective reading patterns in program comprehension.

Busjahn et al. investigated eye movement behavior during source code reading and analyzed the order in which programmers inspect code lines, examining whether reading follows a linear pattern \cite{Busjahn2015}. 
Similarly, Bednarik et al. conducted eye-tracking studies on program comprehension by analyzing fixation durations and distributions across predefined areas of interest in programming tasks \cite{Bednarik2006}. 
However, these studies relied on task-specific code regions or line-based representations and did not abstract gaze behavior into a structural representation that can be applied across different programming tasks. 
In contrast, our study represents gaze movements as transitions between syntactic elements, enabling cross-task structural analysis of gaze patterns associated with comprehension success.

Rodeghero et al. categorized source code into four region types: method declarations, method calls, control flow, and others. 
They analyzed fixation duration for each region while professional programmers summarized source code \cite{Rodeghero2014}.
However, their study focused exclusively on professional programmers and did not compare gaze behavior across groups with different levels of comprehension.
In contrast, our study performs a comparative analysis between participants who successfully completed the comprehension tasks and those who did not. 
Through this comparison, we aim to identify gaze patterns associated with effective program comprehension as well as patterns that are less conducive to successful understanding.

Several tools have been developed to map gaze data to source code elements. 
For example, iTrace provides infrastructure for automatically linking gaze data to fine-grained code elements within IDE environments using tools such as srcML \cite{itrace}.
While such infrastructures enable gaze data to be associated with source code elements, our study focuses on analyzing gaze transition patterns between syntactic elements derived from the syntax tree.

\section{Coordinate-to-syntax mapping method \cite{yoshioka2024}}
Previous work has proposed tools for mapping gaze data to source code elements.
For example, iTrace provides infrastructure that automatically associates gaze data with source code elements within IDE environments using srcML \cite{itrace}.
In this study, we adopt the coordinate-to-syntax mapping method proposed by Yoshioka et al. \cite{yoshioka2024}, which converts coordinate-based gaze data recorded by an eye tracker into transitions between nodes in the syntax tree generated from the source code.

\begin{figure*}[bt]
\begin{center}
\includegraphics[width=0.8\textwidth]{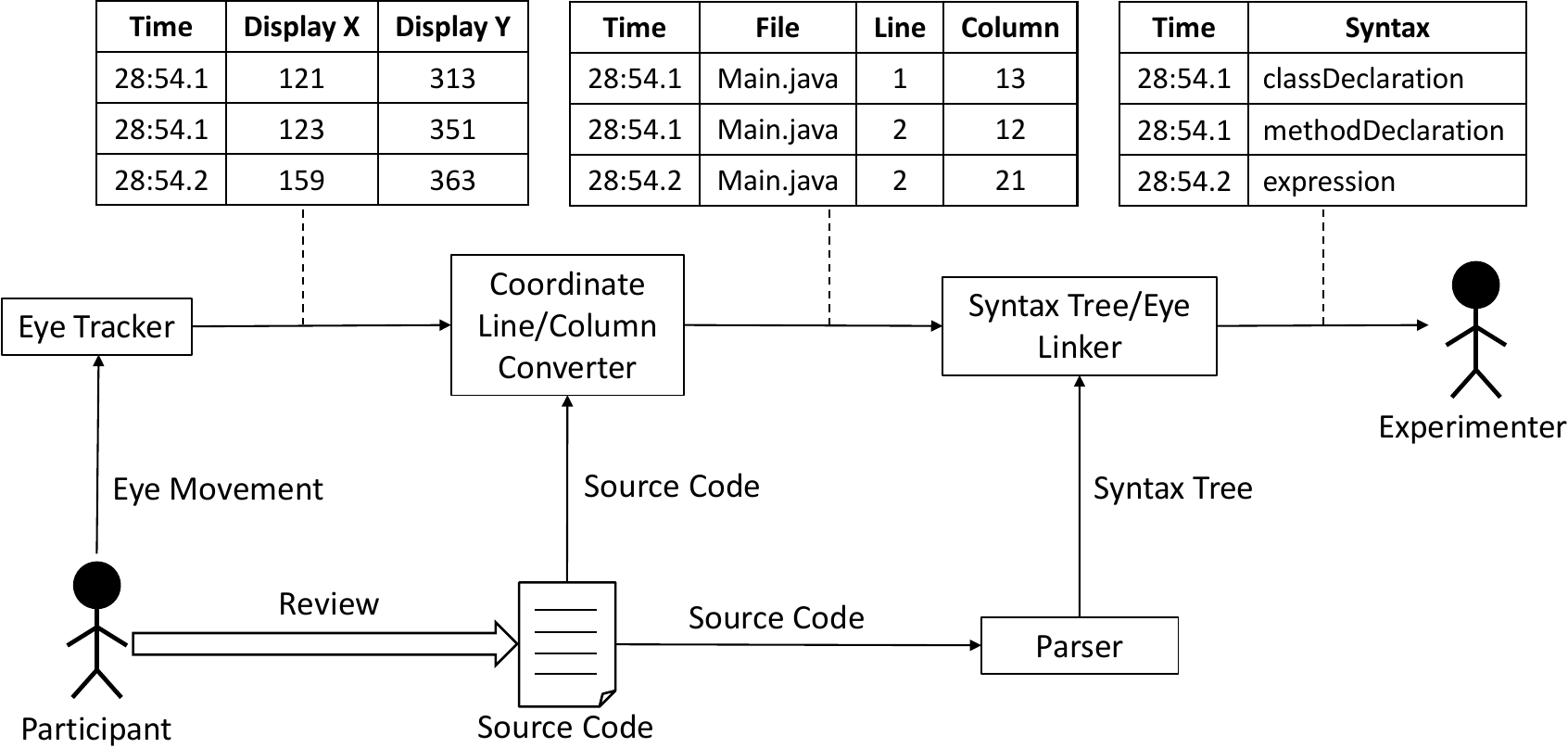}
\caption{Architecture of Yoshioka's Method}
\label{figure:teianshuhou}
\end{center}
\end{figure*}

Figure \ref{figure:teianshuhou} illustrates the architecture of the method. 
In the figure, rectangles represent system modules composing the system, and arrows indicate information flow. 
The eye-tracker records gaze points on the display as time-series coordinate data (e.g., X:121, Y:313). 
The Coordinate Line/Column Converter Module takes these coordinate-based gaze data and the corresponding source code as input, then converts them into source code identifiers and line/column positions (e.g., Main.java, line:1, column:13). 
Based on these line and column values, the module extracts the corresponding word or character from the source code and maps it to a node in the syntax tree obtained through parsing. 
Consecutive gaze on the same word or character is merged into a single fixation to reduce redundancy. 
The Syntax Tree/Eye Linker Module then combines the line/column-based fixation data with the syntax tree generated by the parser and outputs fixation movements at the syntax-node level (e.g., 'classDeclaration'). 
The parser provides structural information for each token, including its line number, column numbers, character length, and syntactic type. 
Using this information, the Linker associates each fixation with the corresponding syntax node, thereby transforming coordinate-level gaze data into node-level transitions on the syntax tree.

\section{Experiment}
\subsection{Task settings}
%Overview
We use the eye-tracking dataset collected in Ishida et al.'s study \cite{ishida2019}.
The experimental design and task settings described below follow those of Ishida et al.~\cite{ishida2019}.
In the original experiment, participants were presented with program specifications written in Japanese and the corresponding Java source code. 
An eye-tracker recorded their gaze movements while they performed program comprehension tasks.
The participants were 14 students (aged 19--21) from the National Institute of Technology (KOSEN), Nara College, who had completed a basic Java programming course.
The experiment was conducted in a quiet room with one participant and two experimenters present. 
The experimental materials were presented on a display. 
Gaze data were collected using a Tobii Eye Tracker 4C, a low-cost (under \$200), non-invasive, screen-based eye tracker with a sampling rate of 90~Hz.

% Tasks
Each participant completed 16 tasks: eight low-difficulty tasks and eight high-difficulty tasks.
Each task consisted of three components: a program specification written in Japanese, the corresponding Java source code, and a comprehension question designed to verify the participant’s understanding of the program behavior (e.g., ``What is the value of 'a' when line 6 is executed for the second time?'').
The authors evaluated the responses against pre-defined correct answers. 
A task was classified as successful if the participant provided the correct answer within the time limit; otherwise, it was classified as unsuccessful, including cases where no answer was provided within the time limit.
Low-difficulty tasks consist solely of a \texttt{main} method and include only single-level loops and conditional branches.
High-difficulty tasks involve multiple methods and/or recursive structures and were designed to be difficult to complete within the time limit.
This design was intended to yield a roughly balanced distribution of successful and unsuccessful cases, enabling meaningful comparison between the two groups.
Table \ref{table:task} summarizes the task specifications and successful/unsuccessful numbers.

\begin{table*}[tb]
\caption{Tasks Used in the Experiment~\cite{ishida2019}. Successful and Unsuccessful indicate the number of participants who correctly answered each task within the time limit and those who did not, respectively.}
\begin{center}
\begin{tabular}{l r l p{5cm} r r r}
\toprule
Difficulty & ID & Task & Specification & LOC & Successful & Unsuccessful \\ \hline
\multirow{8}{*}{Low}
& 1  & Factorial     & Calculate factorial                                    & 12 & 11 & 1 \\
& 2  & SearchMax     & Search for maximum value                               & 12 & 9  & 3 \\
& 3  & PrimeNum      & Prime number check                                     & 15 & 8  & 3 \\
& 4  & SearchMedian  & Search for median value                                & 17 & 11 & 0 \\
& 5  & Power         & Power calculation                                      & 12 & 10 & 2 \\
& 6  & Swap          & Swap two numbers                                       & 12 & 10 & 2 \\
& 7  & Substring     & Check if contains specified substring                  & 23 & 12 & 0 \\
& 8  & ReverseString & Reverse a string                                       & 12 & 13 & 1 \\ \hline
\multirow{8}{*}{High}
& 9  & TowerOfHanoi  & Tower of Hanoi                                         & 24 & 0  & 13 \\
& 10 & NumOfRoute    & Find the number of routes                              & 28 & 3  & 10 \\
& 11 & Permutation   & Enumerate all permutations                             & 37 & 7  & 7 \\
& 12 & Combination   & Find combinations using a recurrence relation          & 25 & 11 & 3 \\
& 13 & PayMoney      & Find coin combinations for payment                     & 28 & 4  & 10 \\
& 14 & StrCombination& Find string combinations                               & 24 & 7  & 6 \\
& 15 & CloudSim      & Cloud movement simulation                              & 25 & 2  & 10 \\
& 16 & lcm\_gcd      & Find least common multiple and greatest common divisor & 15 & 1  & 10 \\ \hline
\end{tabular}
\end{center}
\label{table:task}
\end{table*}

%Other Experimental Settings
The time limit for each task is set to 2 minutes and 30 seconds based on a pilot experiment.
This time constraint was introduced to distinguish between successful and unsuccessful task performance under controlled conditions.
The pilot experiment confirmed that this duration was generally sufficient for low-difficulty tasks but insufficient for high-difficulty tasks.
To control for order effects, the task presentation order was counterbalanced using a Latin square design.

\subsection{Analysis}
To answer RQ1 (Is there a difference in gaze transition patterns related to \texttt{for}-loop structures between successful and unsuccessful participants?), the three syntactic components of a \texttt{for} statement are labeled as \INIT{} (Initialization), \COND{} (Condition), and \UPD{} (Update). 
These elements correspond to the initialization expression, loop continuation condition, and update expression, respectively.
Focusing on gaze movements toward \texttt{for} statements, we extract frequent patterns observed in gaze transitions among the three elements appearing in \texttt{for} statements: \INIT, \COND, and \UPD.
We analyze gaze transition patterns targeting these elements to investigate differences between successful and unsuccessful participants.

To answer RQ2 (Is there a difference in gaze transition patterns across methods between successful and unsuccessful participants?), we analyze elements at a higher level of abstraction.
The proposed method \cite{yoshioka2024} aggregates gaze data into higher-level syntactic elements, allowing transitions to be represented between structural elements such as blocks, methods, and classes.

For both RQs, we apply cSPADE algorithm~\cite{Mohammed2001} to extract frequent sequential patterns. 
The pattern length is set from 1 to 10.
We focus on patterns where the absolute difference in average support between groups is $\ge 0.1$. 
Patterns with support values $> 0.9$ or $< 0.1$ in both groups are excluded, as such patterns represent transitions that are either universally common or extremely rare.

%-------------------------------------------------------------

\section{Results and Discussion}
Out of 256 program comprehension tasks (16 participants $\times$ 16 tasks), 200 tasks were used for analysis after excluding tasks with measurement errors, including 119 successful tasks and 81 unsuccessful tasks.

\subsection{Gaze Transition Patterns at \texttt{for} Loops}
Table \ref{table:for} shows the average support values of gaze transition patterns observed in successful and unsuccessful tasks.
Sequences with a support difference of 0.1 or greater between the successful and unsuccessful groups were extracted as patterns. 
The patterns are sorted in descending order of the support difference, defined as $Successful - Unsuccessful$, where positive values indicate higher support in the successful group.
%The support difference is defined as (Successful - Unsuccessful), where positive values indicate higher support in the successful group.
Significant differences (p < 0.05, chi-squared test) were observed between successful and unsuccessful groups in these gaze transition patterns.
When comparing patterns that appeared more frequently in the successful group with those more frequent in the unsuccessful group, we found that the patterns dominant in the unsuccessful group did not include \UPD. 
This result suggests that participants who successfully completed the tasks may developed a deeper understanding of the \texttt{for} loop by attending to \UPD{} and confirming the loop update process.

% \begin{table}[tb]
% \caption{Transition Patterns at \texttt{for} Loops}
% \centering
% \small
% \begin{tabular}{lclclcl|rrr|rr}
% \hline
% \multicolumn{7}{c|}{Sequence} 
% & \multicolumn{3}{c|}{Average Support Value} 
% & \multicolumn{2}{c}{Chi-Square Test} \\
% \cline{1-12}
% & & & & & & 
% & Successful & Unsuccessful & \DIFF & $\chi^2(1)$ & p-value \\
% \hline

% \UPD & $\rightarrow$ & \COND & & & & 
% & 0.429 & 0.306 & 0.124 & 2460.4 & <0.001 \\

% \UPD & $\rightarrow$ & \COND & $\rightarrow$ & \INIT & & 
% & 0.393 & 0.282 & 0.111 & 2214.3 & <0.001 \\

% \UPD & $\rightarrow$ & \COND & $\rightarrow$ & \INIT & $\rightarrow$ & \COND
% & 0.380 & 0.259 & 0.122 & 2050.3 & <0.001 \\

% \hline

% \INIT & $\rightarrow$ & \UPD & $\rightarrow$ & \COND & $\rightarrow$ & \INIT
% & 0.217 & 0.106 & 0.111 & 738.1 & <0.001 \\

% \hline

% \UPD & $\rightarrow$ & \INIT & $\rightarrow$ & \COND & & 
% & 0.302 & 0.189 & 0.114 & 1722.3 & <0.001 \\

% \hline

% \COND & $\rightarrow$ & \INIT & $\rightarrow$ & \COND & & 
% & 0.793 & 0.944 & -0.152 & 5002 & <0.001 \\

% \hline

% \INIT & $\rightarrow$ & \COND & $\rightarrow$ & \INIT & $\rightarrow$ & \COND
% & 0.776 & 0.944 & -0.169 & 4920.7 & <0.001 \\

% \hline
% \end{tabular}
% \label{table:for}
% \end{table}

\begin{table*}[tb]
\caption{Transition Patterns at \texttt{for} Loops. \INIT{}, \COND{}, and \UPD{} represent the initialization, condition, and update components of a \texttt{for} loop, respectively.}
\centering
\small
\setlength{\tabcolsep}{4pt}
\begin{tabular}{l@{\,$\rightarrow$\,}l@{\,$\rightarrow$\,}l@{\,$\rightarrow$\,}l rrr rr}
\toprule
\multicolumn{4}{c}{\multirow{2}{*}{Sequence}} 
& \multicolumn{3}{c}{Average Support Value} 
& \multicolumn{2}{c}{Chi-Square Test} \\ 
\cline{5-9}
\multicolumn{4}{c}{} 
& Successful & Unsuccessful & Diff & $\chi^2(1)$ & p-value \\
\midrule

\UPD & \multicolumn{3}{@{}l@{}}{\COND}
& 0.429 & 0.306 & 0.124 & 2460.4 & <0.001 \\

\UPD & \COND & \INIT & \COND
& 0.380 & 0.259 & 0.122 & 2050.3 & <0.001 \\

\UPD & \INIT & \multicolumn{2}{@{}l@{}}{\COND}
& 0.302 & 0.189 & 0.114 & 1722.3 & <0.001 \\

\UPD & \COND & \multicolumn{2}{@{}l@{}}{\INIT}
& 0.393 & 0.282 & 0.111 & 2214.3 & <0.001 \\

\INIT & \UPD & \COND & \INIT
& 0.217 & 0.106 & 0.111 & 738.1 & <0.001 \\

\midrule

\COND & \INIT & \multicolumn{2}{@{}l@{}}{\COND}
& 0.793 & 0.944 & -0.152 & 5002.0 & <0.001 \\

\INIT & \COND & \INIT & \COND
& 0.776 & 0.944 & -0.169 & 4920.7 & <0.001 \\

\bottomrule
\end{tabular}
\label{table:for}
\end{table*}

Table~\ref{table:for2} presents the average support values of gaze transition patterns of length two or greater that contain at least one occurrence of each of \INIT, \COND, and \UPD.
No substantial differences in average support values are observed between the successful and unsuccessful groups for these patterns. 
This result indicates that the mere presence of \UPD{} within a gaze transition pattern does not sufficiently explain differences in task success.
Both groups exhibited gaze transitions involving \INIT, \COND, and \UPD; however, the differences appear to lie in the frequency of occurrence, transition structure, and the ways in which these elements are combined.

These findings suggest that participants who successfully completed the tasks repeatedly referred to \UPD{} in relation to other syntactic elements.
In contrast, participants who unsuccessfully completed the tasks, even when briefly attending to \UPD, appear less likely to engage in reciprocal transitions with \INIT{} and \COND, leading to a less integrated understanding of the loop update process.
\\

\noindent\textbf{Answer to RQ1.}
There are differences in gaze transition patterns related to \texttt{for}-loop structures between successful and unsuccessful tasks. 
When understanding for-loops, it is important not only to examine each syntactic element individually but also to attend to transitions that integrate \UPD{} with other elements.

\begin{table}[tb]
\caption{Average Support Values for All Gaze Transition Patterns Contain \INIT, \COND, and \UPD}
\begin{center}
\begin{tabular}{lrrr}
\toprule
\multicolumn{1}{c}{\multirow{2}{*}{Syntactic Element}} 
& \multicolumn{3}{c}{Average Support Value} \\ \cline{2-4} 
\multicolumn{1}{c}{} 
& Successful & Unsuccessful & \DIFF \\ 
\hline
\INIT & 0.224 & 0.225 & -0.002 \\
\COND & 0.227 & 0.229 & -0.002 \\
\UPD  & 0.150 & 0.144 & 0.006 \\
\bottomrule
\end{tabular}
\end{center}
\label{table:for2}
\end{table}

\subsection{Gaze Transition Patterns between Methods}
We analyzed gaze transitions between methods in source code containing multiple methods and performed frequent pattern mining.
Frequent gaze transition patterns were identified for tasks with two methods (Tasks 12, 13, and 14: \texttt{main} and \texttt{method1}) and tasks with three methods (Tasks 9, 10, 11, and 16: \texttt{main}, \texttt{method1}, \texttt{method2}).
In tasks with two methods (Tasks 12–14), both groups frequently alternated their gaze between \texttt{main} and \texttt{method1} (support $\ge 0.89$). 
No significant difference in support values was observed between the successful and unsuccessful groups (difference $\le 0.11$), suggesting that gaze transitions between two methods in simple program structures are not a strong indicator of comprehension success.

Figure~2 illustrates the method call relationships for tasks involving three methods (Tasks 9, 10, 11, and 16). 
Although all tasks involve the same set of method names, the functionality and call structures differ across tasks.
Table~\ref{table:high_abs} shows the average support values of gaze transition patterns observed in these taskGs for successful and unsuccessful tasks, focusing on patterns with a support difference of 0.2 or greater between the two groups.
We present the five patterns with the largest positive differences and the five with the largest negative differences as the most distinctive patterns.
Furthermore, the longest gaze transition pattern within each group was selected as the representative pattern for that group.

\begin{figure}[bt]
\begin{center}
\includegraphics[width=\linewidth]{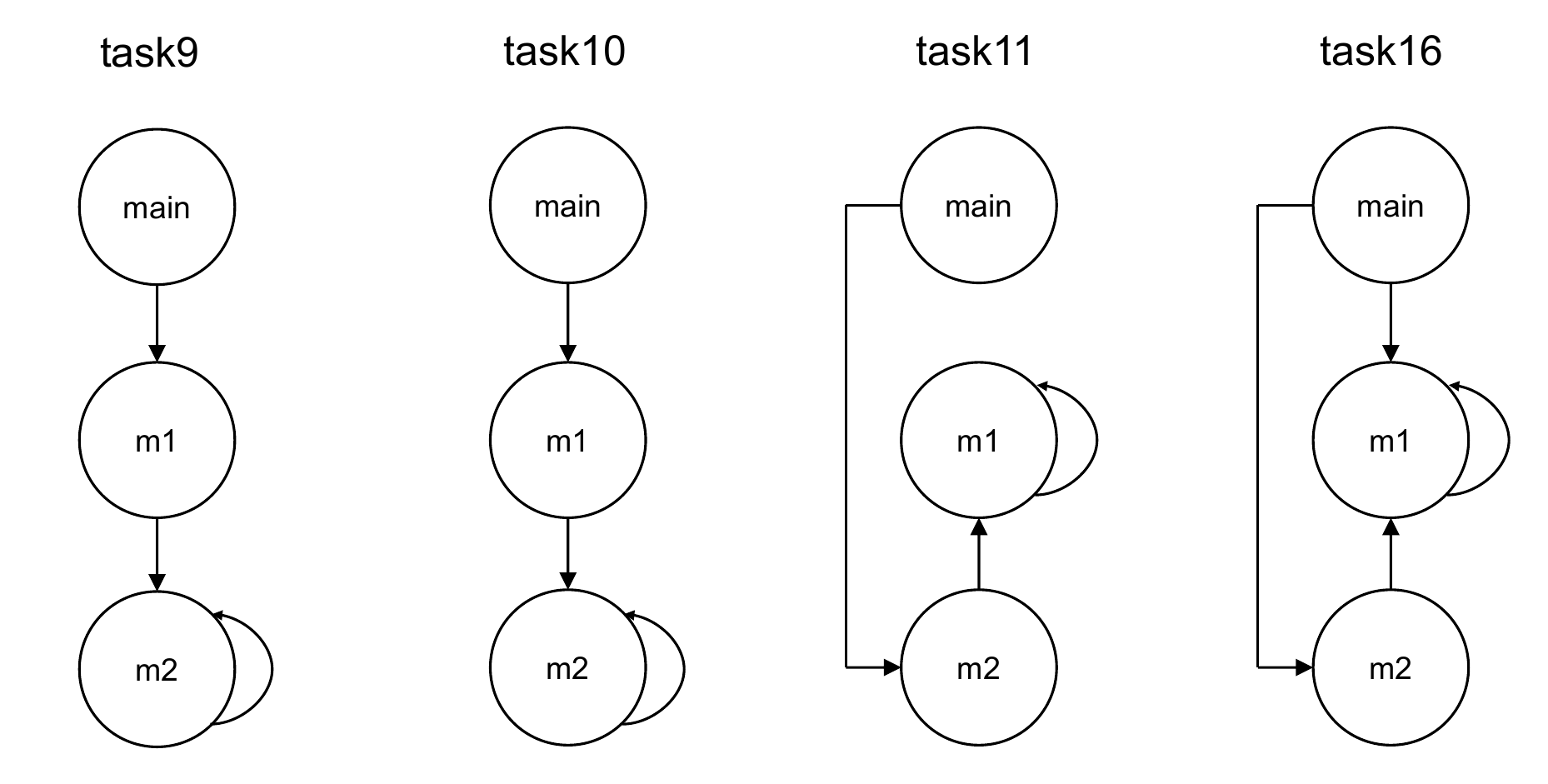}
\caption{Method call relationships for tasks 9, 10, 11, and 16 (m1: method1, m2: method2)}
\label{figure:method3_src}
\end{center}
\end{figure}

\begin{table*}[tb]
\caption{Frequent Patterns Among 3 Methods ($p<0.05$)}
\centering
\small
\setlength{\tabcolsep}{4pt}
\begin{tabular}{l l l@{\,$\rightarrow$\,}l@{\,$\rightarrow$\,}l@{\,$\rightarrow$\,}l@{\,$\rightarrow$\,}l rrr} 
\toprule
\multicolumn{1}{c}{\multirow{2}{*}{Group}} & \multirow{2}{*}{ID} & \multicolumn{5}{l}{\multirow{2}{*}{Sequence}}   & \multicolumn{3}{c}{Average Support Value} \\ \cline{8-10} 
\multicolumn{1}{c}{}                       &                     & \multicolumn{5}{l}{}                            & Successful    & Unsuccessful   & Diff    \\  
\midrule

\multirow{5}{*}{Successful}
&$S_1$  & method1 & main    & method1 & main    & method2 & 0.381 & 0.130 & 0.251 \\
&$S_2$  & method2 & method1 & method2 & method1 & main    & 0.841 & 0.536 & 0.305 \\
&$S_3$  & main    & method2 & method1 & main    & method1 & 0.476 & 0.155 & 0.321 \\
&$S_4$  & method1 & main    & method2 & method1 & main    & 0.381 & 0.050 & 0.331 \\
&$S_5$  & main    & method1 & main    & method2 & method1 & 0.444 & 0.099 & 0.345 \\

\midrule

\multirow{5}{*}{Unsuccessful}
&$U_1$    & main    & method1 & main   & method1 & main    & 0.254 & 0.572 & -0.318 \\
&$U_2$    & method2 & main    & method1 & main   & method1 & 0.048 & 0.357 & -0.310 \\
&$U_3$    & main    & method1 & method2 & main    & method1 & 0.111 & 0.421 & -0.309 \\
&$U_4$    & method1 & method2 & main    & method1 & main    & 0.048 & 0.337 & -0.289 \\
&$U_5$    & method2 & method1 & method2 & method1 & method2 & 0.476 & 0.699 & -0.222 \\

\bottomrule
\end{tabular}
\label{table:high_abs}
\end{table*}

The results show that gaze transition patterns of participants who successfully completed the tasks include the transition \texttt{main} $\rightarrow$ \texttt{method2} (ID $S_1$, $S_3$, $S_4$, and $S_5$), whereas this transition was not observed among unsuccessfully completed tasks.
The average support value for the \texttt{main} $\rightarrow$ \texttt{method2} transition was higher for successful tasks (0.889) than for unsuccessful tasks (0.665), and the difference was statistically significant ($p \textless 0.01$ chi-squared test).
In Tasks 11 and 16, \texttt{method2} is called from the \texttt{main} method.
Therefore, \texttt{main} $\rightarrow$ \texttt{method2} transition likely reflects gaze behavior that follows the method call relationship.
Other gaze transitions consistent with method calls include \texttt{main} $\rightarrow$ \texttt{method1}, \texttt{method1} $\rightarrow$ \texttt{method2}, and \texttt{method2} $\rightarrow$ \texttt{method1}. 
These transitions occur between adjacent methods and appear frequently across many gaze patterns, suggesting that they represent common reading behavior when inspecting source code.

In addition, the patterns  (ID $S_2$, $S_3$, and $S_4$) of Table~\ref{table:high_abs} include the transition \texttt{method2} $\rightarrow$ \texttt{method1} $\rightarrow$ \texttt{main}, which was not observed among participants who unsuccessfully completed the tasks.
In Tasks 9 and 10, the execution flow involves calling \texttt{method1} from \texttt{main} and subsequently calling \texttt{method2} from \texttt{method1}.
Therefore, the transition \texttt{method2} $\rightarrow$ \texttt{method1} $\rightarrow$ \texttt{main} represents gaze movement in the direction opposite to the execution flow.
This suggests that successful participants traced the execution flow backward to verify argument passing and data flow.
Furthermore, among the patterns that differed between successful and unsuccessful participants, transitions that repeatedly alternate between two methods (e.g., \texttt{main} $\rightarrow$ \texttt{method1} $\rightarrow$ \texttt{main}) were rarely observed.
The results for tasks with two methods and the patterns shown in Table \ref{table:high_abs} suggest that simple alternation between two methods has limited influence on successful program comprehension.
%These findings address RQ2 (Is there a difference in gaze transition patterns across methods between successful and unsuccessful participants?).
\\

\noindent\textbf{Answer to RQ2}\\
There are differences in gaze transition patterns between successful and unsuccessful program participants when multiple methods are involved. 
Successful participants tend to follow method call relationships and, in some cases, trace execution flow backward to confirm data flow. 
Such structured gaze transitions appear to support deeper understanding of program behavior.

\section{Conclusion}
In this study, gaze data recorded during program comprehension were converted into transitions between syntactic nodes, and frequent pattern mining (cSPADE) was applied to analyze gaze transition patterns.
In the experiment, we compared gaze data obtained during program comprehension tasks, focusing on gaze transition patterns related to \texttt{for}-loop structure and those across multiple methods, between successful and unsuccessful tasks. 
Through this analysis, we identified frequently observed gaze transition patterns associated with comprehension outcomes.
The analysis of gaze transition patterns related to \texttt{for} statements revealed differences in how participants inspected \texttt{for} statements. 
The results suggest that attending to  all three elements, \INIT{}, \COND{}, and \UPD{}, may support more structured understanding of loop behavior.
The analysis of gaze transition patterns across multiple methods also revealed differences between successful and unsuccessful participants.
In particular, task success appears to be associated with following method call relationships and, in some cases, tracing data flow across method boundaries.

As future work, we plan to conduct more fine-grained analyses of gaze transition patterns related to control flow.
In particular, we will extend the analysis to four elements of a \texttt{for} statement: \INIT, \COND, \UPD, and the loop body that is repeatedly executed.
Previous analyses focusing only on the three header elements mainly captured how participants interpreted the operation of the \texttt{for} statement. 
By additionally incorporating the loop body, we aim to investigate how readers shift their gaze between the loop body and the elements \INIT, \COND, and \UPD. 
This extended analysis may help reveal more effective reading strategies for understanding loop behavior as a whole.

\newpage

\bibliographystyle{ACM-Reference-Format}
\nocite{*}
\bibliography{literature}
\end{document}